\documentclass[preprint]{aastex}
\usepackage{emulateapj5}
\input psfig.tex

\shorttitle{EARLY TYPE GALAXY PROGENITORS}
\shortauthors{MCCARTHY ET AL.}

\begin{document}

\title{ The Las Campanas IR Survey: \\ 
   Early Type Galaxy Progenitors Beyond Redshift One}

\author{
P.~J.~McCarthy\altaffilmark{1}, 
R.~G.~Carlberg\altaffilmark{2}, 
H.-W.~Chen\altaffilmark{1}, 
R.~O.~Marzke\altaffilmark{1,3}, 
A.~E.~Firth\altaffilmark{4}, 
R.~S.~Ellis\altaffilmark{5}, 
S.~E.~Persson\altaffilmark{1}, 
R.~G. ~McMahon\altaffilmark{4}, 
O.~Lahav\altaffilmark{4}, 
J.~Wilson\altaffilmark{1},
P.~Martini\altaffilmark{1},
R.~G.~Abraham\altaffilmark{2}, 
C.~N.~Sabbey\altaffilmark{4},
A.~Oemler\altaffilmark{1}, 
D.~C.~Murphy\altaffilmark{1},
R.~S.~Somerville\altaffilmark{4}
M.~G.~Beckett\altaffilmark{1,4},
J.~R.~Lewis\altaffilmark{4},
C.~D.~MacKay\altaffilmark{4}
}

\submitted{Accepted for publication in the Astrophysical Journal $Letters$}

\altaffiltext{1}{Carnegie Observatories, 813 Santa Barbara St,
        Pasadena, CA 91101}
\altaffiltext{2}{Department of Astronomy, University of Toronto,
        Toronto ON, M5S~3H8 Canada}
\altaffiltext{3}{Department of Astronomy and Physics, San Francisco State University,
        San Francisco, CA }
\altaffiltext{4}{Institute of Astronomy, Cambridge CB3 OHA, UK}
\altaffiltext{5}{Department of Astronomy, Caltech 105-24,
        Pasadena, CA 91125}

\begin{abstract}
   
We have identified a population of faint red galaxies from a 0.62
square degree region of the Las Campanas Infrared Survey whose
properties are consistent with their being the progenitors of early-type 
galaxies.  The optical and IR colors, number-magnitude
relation and angular clustering together indicate modest evolution
and increased star formation rates among the early-type field population 
at redshifts between one and two.  The counts of red galaxies
with $H$ magnitudes between 17 and 20 rise with a slope that is
much steeper than that of the total $H$ sample.  The surface density of red
galaxies drops from roughly 3000 per square degree at $H = 20.5, I-H > 3$
to $\sim 20$ per square degree at $H = 20, I-H > 5$.  The $V-I$ colors
are approximately 1.5 magnitudes bluer on average than a pure old
population and span a range of more than three magnitudes.  The strength of
the angular clustering of the red galaxies is an order of magnitude
larger than that of the full galaxy sample.  The colors, and
photometric redshifts derived from them, indicate that the red
galaxies have redshift distributions adequately described by
Gaussians with $\sigma_z\simeq 0.2$ centered near redshift one, with
the exception that galaxies having $V-I<1.6$ and $I-H>3$ are primarily
in the $1.5\la z \la 2$ range.  We invert the angular correlation
functions using these n($z$) and find co-moving correlation lengths of
$r_0\simeq 9-10 h^{-1}$Mpc at $z\simeq 1$, comparable to,
or larger than, those found for early-type galaxies at lower
redshifts. A simple photometric evolution model reproduces the counts
of the red galaxies, with only a $\sim 30$\% decline in the underlying space
density of early-type galaxies at $z \sim 1.2$. The colors indicate 
characteristic star formation
rates of $\sim 1M_{\odot}/yr$ per $10^{10}M_{\odot}$.  We suggest on the basis of the
colors, counts, and clustering that these red galaxies are the bulk of
the progenitors of present day early-type galaxies. 

\end{abstract}


\section{Introduction}

The earliest deep images at high latitude obtained with near-IR arrays
revealed a population of galaxies not represented in optical surveys
(e.g. Elston, Rieke, \& Rieke 1988; McCarthy, Persson \& West 1992; Hu
\& Ridgway 1994).  These objects have colors and apparent magnitudes
close to those expected of evolved massive galaxies at $1 < z < 3$.
Selections based on $R - K$ lead to a heterogeneous population
containing evolved stellar populations (Spinrad et al. 1997; Soifer et
al. 1999), reddened objects (Graham \& Dey 1996; Cimatti et al. 1998;
Dey et al. 1999, Barger et al. 2000), and cool stars.  Redshifts for
either dust-free or highly reddened objects are primarily in the $1 <
z < 2$ range (Graham \& Dey 1996; Soifer et al. 1999; Liu et al. 2000,
Cowie et al. 2001).  While the reddest examples (Smail et al
1999) are sometimes associated with bright sub-mm sources and 
cm-wave sources, the bulk of the red galaxies do not appear to be strong cm to
sub-mm continuum sources. 

A key issue in the hierarchical assembly picture of galaxy formation
is the identification of the redshift range over which the most
massive galaxies assemble.  The effects of merging at $z < 1$ are
minor: the luminosity function evolves primarily through passive
evolution (Lilly et al. 1995, Cowie et al. 1996) and the merger rate
is low (Carlberg et al 2000). The rate of mass buildup via merging is
dependent on the density of galaxies, the correlation function at
relevant separations, and the distribution of pairwise velocities. For
major mergers to play a dominant role in galaxy building, the
clustering of at least some component of the normal galaxy population
must be large relative to that at the current epoch.  Near-IR
selection provides a powerful tool for removing the foreground of
faint low redshift star forming galaxies and offers a path toward
selecting galaxies at $z \sim 1$ by stellar mass.

The Las Campanas Infrared (LCIR) survey was crafted to select
early-type galaxies at $1 < z < 2$ (Marzke et al. 1999).
It is a near-IR survey to $H \sim 21$ over a significant
fraction of a square degree, supplemented by photometry in the UBVRIz$^{\prime}$
bands.  In this {\em Letter} we present the basic observational
results for the galaxy counts, colors and clustering properties, along
with their implications based on simple modeling. Future papers will
augment these data and refine and broaden the interpretation.

\section{Observations \& Results} 

\subsection{Observations}

The data presented here are drawn from a subset of the LCIR survey
covering an area of 0.62 square degrees to a depth ranging from H$=
20$ to 21 in four fields: Hubble Deep Field South (HDFS), Chandra Deep
Field South (CDFS), SSA22 and the NTT Deep Field. 
The $H$ data, obtained with the CIRSI camera (Beckett et al. 1998), are
complemented by deep optical imaging: VRI in all fields, U \& B in
HDFS, and z$^{\prime}$ in CDFS. The details of the near-IR survey are
described in Chen et al. (2001$a$) and Firth et al. (2001),
while the optical data are discussed in Marzke et al. (in prep).
Photometric catalogs were extracted for each field using $4^{''}$
aperture Johnson magnitudes.

\psfig{file=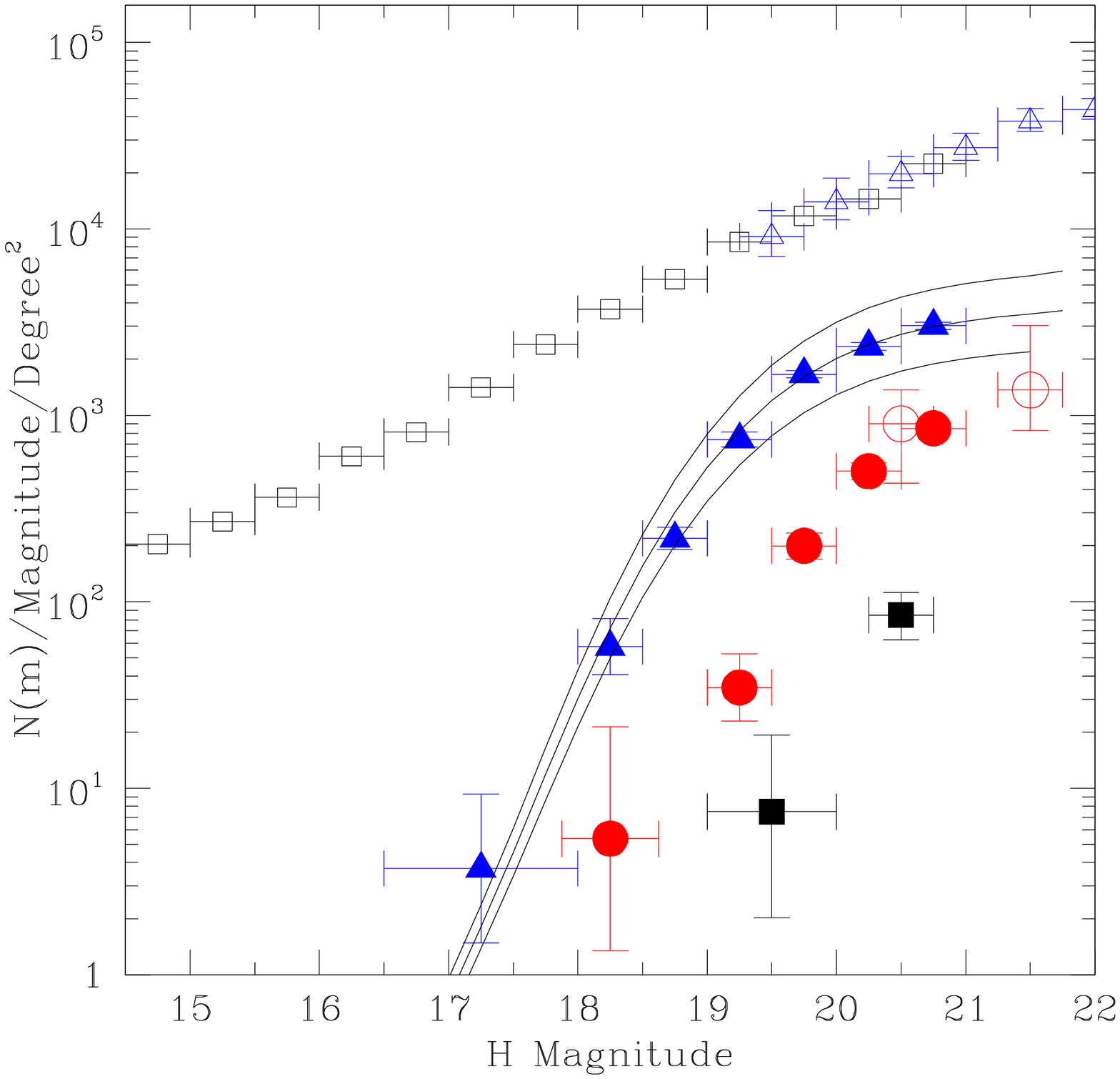,width=8.5cm,angle=0}
\figcaption[]{ The total and red galaxy counts as a function of $H$ magnitude over 0.62
square degrees in four fields.  The open squares represent the
complete galaxy sample, after removal of stars.
The open triangles are the NICMOS based $H$ counts from Yan
et al. (1997). The filled triangles and circles are the H
counts for $I-H>3$ and 4, respectively.  The error bars show the 95\%
confidence interval assuming Poisson statistics.  The open circles at
$H = 20.5$ and 21.5 are from the $R-H>5$ NICMOS sample of Yan et al.
(2000). The filled squares are the counts for $I-H>5$ from the CDFS and HDFS fields only.
The curves show the counts for $I-H>3$ derived from the
evolving population models described in the text with $p = 0, 0.5,$ and 1.
The characteristic redshift for the $I - H >3$ model is 1.2.
\label{fig1}}
\vskip 0.3cm

\subsection{ N(m) for Red Galaxies }

  In Figure 1 we present the differential number magnitude relations
for the complete $H$ selected sample and $I - H$ color selected
sub-samples from the HDFS, CDFS, SSA22 and NTT fields.  These counts are
based on $4^{''}$ aperture magnitudes with aperture and incompleteness
corrections derived from point sources. Each field was used only to its
95\% completeness limit.  A full analysis of the number counts and their completeness are given 
in Chen et al. (2001$a$) and Firth et al. (2001). The complete $H$
counts are similar in slope, dlog(n)/dm, and amplitude to the deep $H$
counts derived by Yan et al. (1997).  The $I - H> 3$ counts are quite
steep, dlog(n)/dm$ = 1.1$ in the $17 <$ H $< 19.5$ range.  At fainter
levels the counts must flatten and the slope of the $I - H>3$ subsample appears
to change near $H =19.5$.  The counts in the redder color bin
appear to have the same slope as the $I - H>3$
counts for $H<20$ but with surface densities that fall roughly one
order of magnitude per magnitude of increasingly red color. At bright
magnitudes (e.g. H$\sim 17.5$) the red galaxies constitute roughly
0.5\% of the total $H$ selected population with a surface density of a
few per square degree, while at H$ = 20.5$ they contribute roughly
10\% of the total population at a density of $\sim 3000$ per square
degree. We note that our primary color-cut, $I-H > 3$, is roughly one magnitude
bluer than the canonical ``ERO'' color threshold of $R - K >6$.
 
  We use a set of simple evolutionary models to examine both the
red counts and the colors (\S\ 2.3), computed with the PEGASE.2 code
(Fioc \& Rocca-Volmerange 1997) and the Gardner et al. (1997) local K-band
luminosity function.  We identify the
red galaxy population as those galaxies brighter than $M_\ast-1$ as fitted
with a Schechter function with $\alpha=1$ and $M_\ast({\rm
red})=M_\ast-0.2$ and $\phi_\ast({\rm red})=0.15\phi_\ast$.  Assuming
the $I-H$ colors at the faint limit of our sample have a precision of
$0.25$ magnitudes, this model luminosity function is evolved using a $\tau = 1$
decaying star formation rate model and a number density that evolves as
$(1 + z)^{-p}$ in a $\Omega_M=0.3,
\Omega_\Lambda=0.7$ cosmology to predict the red galaxy number counts 
in Figure~1. Density evolution with $p = 0.5 \pm 0.2$ agrees well with the
observed counts of galaxies with $I - H > 3$, and implies a decline in
true space density of $\sim 30$\% at $z = 1.2$.

There have been a variety of measurements of the surface density of
red objects from near-IR surveys, primarily based on R - K selected
samples. At the faint end our surface density of 1000 per square
degree for $I-H>4$ agrees well with the NICMOS measurements reported
by Yan et al. (2000) for $R - H >5$.  Our counts at $H = 20, I - H>4$
agree well with the surface density of 150 per square degree for $R -
K > 6, K < 19.0$ reported by Thompson et al. (1999) and our counts at 
$I - H>3$ agree with those reported by Daddi et al. (2000) for $R - K>5$.

\subsection{ Optical and Near-IR Colors  }

Stars and galaxies at different redshifts have distinctive signatures
in an optical near-IR color-color diagram. In Figure 2 we plot the
$V-I$ vs.  $I-H$ diagram from the CDFS field, for $19\le H
\le 20.5$. The points are color-coded on the basis of 
photometric redshifts (Chen et al. 2001$b$; Firth et al. 2001). The
red galaxies with $I-H>3$ span more than three magnitudes in
$V-I$. This wide range of optical colors worked against previous
searches for early-type galaxies at $z >1$ that required colors that
matched purely passive models. It is particularly notable that very
few of the red galaxies have $V-I\sim 3$, the color expected of purely
old stellar populations. Our estimated 50\% completeness limit in $V$
is slightly beyond 26 magnitudes, so we should be missing very few of these
inactive red galaxies. Few such objects are seen in deeper fields (e.g. Moustakas et al. 1997)
We note that our survey area is such that we only sample 
the field population and expect to detect few, if any, rich clusters.

Superimposed on the data in Figure 2 are the loci of four evolutionary
models spanning the $0 < z < 2$ range computed using the PEGASE.2 code
(Fioc \& Rocca-Volmerange 1997).  Note that virtually all objects to
the left of the models are stars. With increasing
redshift the colors initially become redder, but as the $V$ band
begins to sample the rest UV part of the spectrum beyond $z = 1$
it begins to brighten.  The model that best matches the red galaxy
population, an exponentially declining star formation rate with $\tau
= 1$Gy, is the same model that reproduces the red counts shown in
Figure 1. At the redshift of observation the star-formation rates are
not insignificant, the rate being $\sim 1 M_{\odot}/yr$ per $10^{10}M_{\odot}$,
comparable to the star formation rates in the general field galaxy population at comparable redshifts.
The $I-H$ defined red galaxies with the bluest $V-I$ colors
are expected to lie at $1.5 < z < 2$ on the basis of this model. The
photometric redshifts support this as shown by the significant number
of $z_{ph} > 1.5$ objects in the lower right portion of the color-color
plane.

\psfig{file=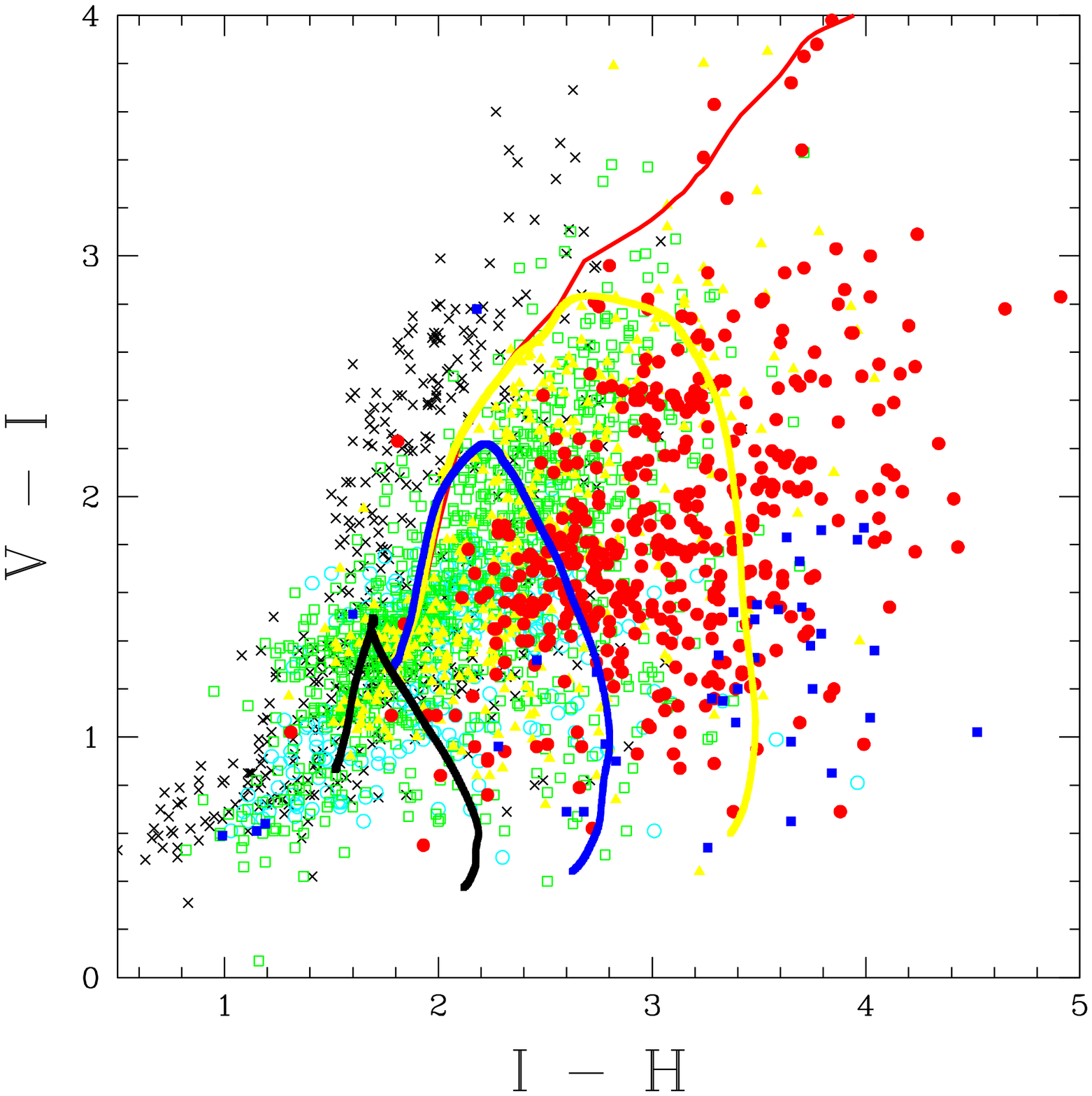,width=8.5cm,angle=0}
\figcaption[]{ The $V-I$ vs. $I-H$ colors for objects with $19 < H < 20.5$ in the
CDFS field.  The points are color coded on the basis of their
photometric redshift: stars (black, crosses), $z<0.25$ (cyan, open circles), $0.25<z<0.75$
(green, open squares), $0.75<z<1.0$ (yellow, filled triangles),
$1<z<1.5$ (red, filled circles), and $1.5<z<2.0$
(blue, filled squares).  The model curves show the loci of evolving population models
with various star formation laws: a single burst with $z_f = 30$
(red), exponentially decline with $\tau = 1 \& 2$ Gyr (yellow and
blue), and continuous (black).  The yellow curve is from the same
model used in Figure 1.
\label{fig2}}
\vskip 0.3cm

\subsection{Angular Clustering}

   We have computed the angular correlation function of the red
objects in our fields. The deepest and most complete data sets come
from the HDFS and CDFS survey areas. The CDFS field covers 561
$\Box^{'}$ with an average 90\% completeness $H$ depth of 20.5, The
HDFS field covers 847 $\Box^{'}$ to an average 90\% completeness
depth of 20.1. We use the Landy \& Szalay (1993) algorithm for
estimating $\omega(\theta)$.  Uncertainties are computed from the
differences between the two fields.  We use the population of stars
and faint blue galaxies as the random distribution. As a check we determined 
$\omega(\theta)$ for a complete $H$-selected sample using a
generated random sample and a completeness map for a sub-region of the
HDFS field and recover a clustering angle of approximately $2^{''}$.

In Figure 3 we display the angular correlation function for the full
H-selected population and for the red galaxies in three apparent
magnitude bins, $18 < H < 19$, $19 < H < 20$, and $20 < H < 20.5$. The
correlation angles, $\theta_0$, derived from fits to
$\omega(\theta)=(\theta_0/\theta)^{\gamma-1}$ ($\gamma=1.8$ is
assumed), are given in Table~1. The clustering shows very steep
color and magnitude dependencies, allowing a strong test of
the estimated redshift distribution and the spatial correlation
length.  There are two primary trends: the red galaxies are very much
more correlated than the entire $H$ limited sample, and, the angular
clustering increases much faster with decreasing limiting magnitude
than can be expected on the basis of depth in a uniformly distributed
population. The natural interpretation is that the red galaxies are
distributed in a relatively narrow redshift shell around $z = 1$.
In Table~1 we report the dependence of clustering on color in various
sub-samples and find that the clustering increases for progressively
redder sub-samples. We split the $I-H>3$ galaxies into sub-samples with
red and blue rest-frame UV colors at $V-I=1.8$, finding that both 
sub-samples have increased correlations compared to the whole, and hence 
occupy narrower redshift ranges.

\subsection{Spatial Clustering}

The red galaxy counts, colors, and angular clustering together allow
us to infer their statistical distribution in redshift. Here we
present a simple population model approach using the evolving luminosity
function as described in \S\ 2.2.

To invert the $\theta_0$ into co-moving correlation length, $r_0(z)$, we use the relativistic
generalization of Limber's equation for $\xi(r,z)=(r_0(z)/r)^\gamma$,
\begin{equation}
\omega(\theta) = A(\gamma)
\theta^{1-\gamma} N^{-2}
\int n^2(z) r_0^\gamma(z) x^{1-\gamma} \, {{H(z)}\over{c}} dz,
\end{equation}
where $N=\int n(z)\, dz$,
$A(\gamma)=\Gamma(\onehalf)\Gamma((\gamma-1)/2)/ {\Gamma(\gamma/2)}$,
and $H(z) = H_0
[\Omega_M(1+z)^3+\Omega_R(1+z)^2+\Omega_\Lambda]^{1/2}$, with
$\Omega_M+\Omega_R+\Omega_\Lambda=1$. The co-moving distance, $x(z)$,
is computed for our adopted cosmology $\Omega_M=0.3,
\Omega_\Lambda=0.7$.

We approximate the $n(z)$ distributions with Gaussians fit to the results of the
population model discussed in \S\ 2.2 (see Table 1).
Determinations of $n(z)$ from our photometric redshifts in the CDFS field yield
mean redshifts and characteristic widths for the $I - H>3$ and $I - H
> 3.5$ sub-samples quite close to those listed
in Table 1.  An additional constraint comes from redshifts for objects in similar color and magnitude 
ranges from the Caltech Faint
Galaxy Redshift Survey (Cohen et al. 1999).  The Cohen et al. sample
is $K$ selected, but if we adopt $H - K = 1$ for the red
population, we can compare the redshift distribution of their $K = 20$
sample. A red color cut yields $<z> = 1.2$, $\sigma_{z} \sim 0.15$,
a distribution somewhat narrower than that derived from the
photometric redshifts, as expected. 
We use our population model to mitigate the small number statistics in the
photometric redshifts, particularly for the reddest colors and brightest magnitudes.
We adopt $<z> = 1.2$ for the red galaxies, 0.9 for the $18\le H \le 19$ range and 1.6 for the
$V-I<1.8$ blue sub-sample of the $I-H$ red galaxies.

The inferred co-moving correlation lengths, $r_o$, are listed in
Table~1. The correlation lengths increase as $\sigma_z^{0.55}$ and
have only a very weak dependence on the mean redshift. Although these
correlation lengths are large, they are comparable to those found for
early-type galaxy populations at low redshifts (Davis \& Geller 1976, Guzzo et
al. 1997, Willmer, da Costa \& Pellegrini 1998). Our larger inferred values
of $r_0$ at high redshift compared to local early-types could be
the result of an increasing biasing with redshift, as
expected theoretically (Mo \& White 1996), or simply an overestimate
of the width of the redshift distribution.

\section{Conclusions}

We have identified a large sample of faint red galaxies from a 0.62 square degree
area of the Las Campanas IR Survey. While the present data set does not allow a direct constraint on the
contribution from heavily reddening star forming galaxies, 
the counts, colors and clustering statistics of these galaxies
are all consistent with their being a mildly evolved progenitor of the present day
early-type field population. We find evidence for a modest change in
the co-moving space density of early-type galaxies to $z \sim 1.2$. The 
star formation rates inferred from the rest-frame UV
colors suggests that the objects are largely, but not
completely, assembled by the epoch of observation.  If the red galaxies are predominantly
old stellar systems, their numbers are larger than predicted in galaxy evolution
models with strong merging (e. g.  Kauffmann \& Charlot 1998,
Somerville et al 2001).  The use of a color selection that does not
extend to the rest-frame UV allows us to recognize nearly-passively
evolving objects with greater confidence than selections spanning large color baselines.

\psfig{file=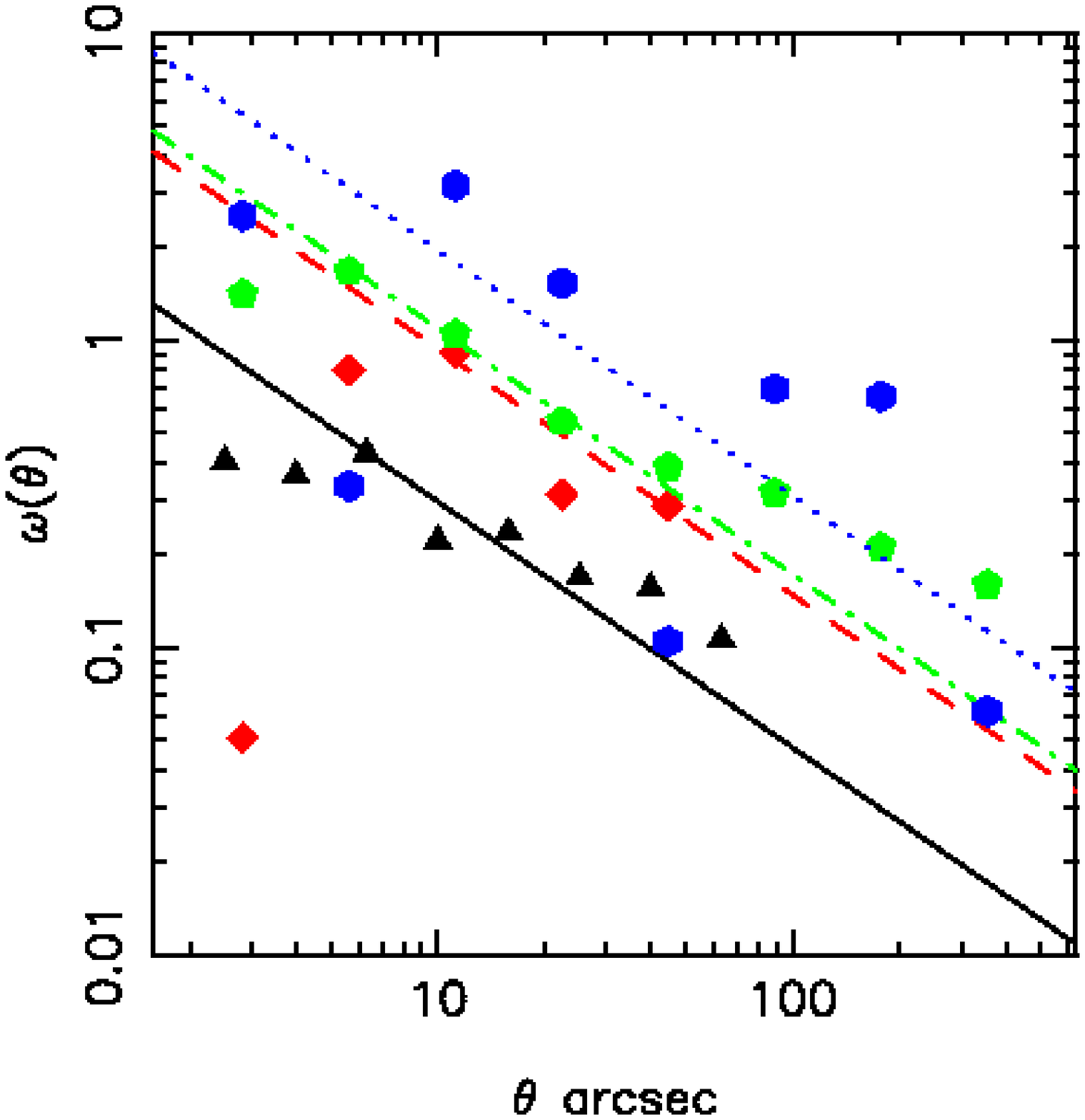,width=8.5cm,angle=0}
\figcaption[]{The angular correlation function for various color and magnitude sub-samples:
all colors for $19 < H < 20.5$ (black triangles), and $I-H>3$ samples with
$18\le H \le 19$(blue circles), $19\le H \le 20$ (green pentagons), and $20\le H \le 20.5$ (red diamonds).
The strong increase with increasing brightness is primarily the result of a narrowing redshift
distribution. In all cases $r_0 \sim 9h^{-1}$ Mpc.
\label{fig3}}
\vskip 0.3cm

\section{Acknowledgments}
  
    This research was supported by the National Science Foundation
under grant AST-9900806 and NSERC of Canada.  The CIRSI camera was
made possible by the generous support of the Raymond and Beverly
Sackler Foundation. 

\clearpage
\eject

\begin{table}
\begin{center}
\begin{tabular}{lcrcrr}
\multicolumn{6}{c}{Table 1. Clustering Properties of Red Galaxies} \\
\hline
\hline
\multicolumn{2}{c}{Subsample}  &\multicolumn{1}{c}{$\theta_0$} & & 
& \multicolumn{1}{c}{$r_0$} \\
\cline{1-2}
\multicolumn{1}{c}{$H$} & \multicolumn{1}{c}{$I-H$} & 
\multicolumn{1}{c}{($''$)}  & 
\multicolumn{1}{c}{$<z>$}& \multicolumn{1}{c}{$\sigma_z$} & \multicolumn{1}{c}{$h^{-1}$Mpc} \\
\hline
 $20.0 - 20.5$ & $3.0 - 5.0$ & $5.0\pm 0.2$           & 1.2& 0.30 & 8.4   \\
 $19.0 - 20.0$ & $3.0 - 5.0$ & $15.0\pm 1.0$          & 1.2& 0.15 & 9.5   \\
 $18.0 - 19.0$ & $3.0 - 5.0$ & $50.0\pm 7.0$          & 0.9& 0.07 & 10.7  \\
 $20.0 - 20.5$ &  ALL        & $2.0\pm 0.2$           & 0.7& 0.30 & 5.7   \\
 $19.0 - 20.5$ & $3.0 - 5.0$ & $6.7\pm 0.4$           & 1.2& 0.30 & 9.8   \\
 $19.0 - 20.5$ & $3.5 - 5.0$ & $6.3\pm 0.5$           & 1.2& 0.30 & 9.5   \\
 $19.0 - 20.5$ & $4.0 - 5.0$ & $11\pm0.5$             & 1.2& 0.20 & 9.7   \\
 $19.0 - 20.5^\dagger$ & $3.0 - 5.0$ & $15\pm2$       & 1.2& 0.15 & 9.6   \\
 $19.0 - 20.5^{\dagger\dagger}$ & $3.0 - 5.0$ & $12\pm1$       & 1.6& 0.2  & 9.8   \\
\hline
\multicolumn{6}{l}{$\dagger V - I >1.8,~~ \dagger\dagger V - I < 1.8$.} \\
\end{tabular}
\end{center}
\end{table}
\end{document}